# phrosty: A difference imaging pipeline for Roman

**Lauren N. Aldoroty**[1,¶], **Lei Hu**[2], **Rob A. Knop**[3], **Cole Meldorf**[6], **Daniel Scolnic**[1], **Shu Liu**[4], **W. Michael Wood-Vasey**[4], **Marcus Manos**[5], **Lucas Erlandson**[5], **Rebekah Hounsell**[7,8], **Ben Rose**[9], **Masao Sako**[6], **Michael Troxel**[1], and **The Roman Supernova Cosmology Project Infrastructure Team**[1]

**1** Department of Physics, Duke University, Durham, NC 27708, USA  **2** Department of Physics, Carnegie Mellon University, Pittsburgh, PA 15213, USA  **3** Lawrence Berkeley National Laboratory, 1 Cyclotron Road, MS 50B-4206, Berkeley, CA 94720, USA  **4** University of Pittsburgh  **5** NVIDIA, 2788 San Tomas Expressway, Santa Clara, CA 95051  **6** Department of Physics and Astronomy, University of Pennsylvania, 209 South 33rd Street, Philadelphia, PA 19104, USA  **7** University of Maryland, Baltimore County, Baltimore, MD 21250, USA  **8** NASA Goddard Space Flight Center, Greenbelt, MD 20771, USA  **9** Department of Physics and Astronomy, Baylor University, One Bear Place 97316, Waco, TX 76798-7316, USA  **¶** Corresponding author

## Summary

NASA's Nancy Grace Roman Space Telescope (*Roman*) will provide an opportunity to study dark energy with unprecedented precision using several techniques, including measurements of Type Ia Supernovae (SNe Ia). Here, we present `phrosty` (PHotometry for ROman with SFFT for tYpe Ia supernovae): a difference imaging pipeline for measuring the brightness of transient point sources in the sky, primarily SNe Ia, using *Roman* data. `phrosty` is written in Python. We implement a GPU-accelerated version of the Saccadic Fast Fourier Transform (SFFT) method for difference imaging.

## Statement of need

The Nancy Grace Roman Space Telescope (*Roman*) is NASA's next flagship mission, designed for near-infrared (NIR) observations (Akeson et al., 2019; Spergel et al., 2013, 2015). There will be several core community surveys that focus on different scientific goals. The High Latitude Time Domain Survey (HLTDS) is one such core survey and is optimized for transient astronomy, which is the study of astronomical objects that change over time. One of the survey's key goals is to collect data for a cosmological analysis with Type Ia Supernovae (SNe Ia). The core component of this survey will last two years, and have a cumulative 3792 hours of exposure time. The survey will yield an average of 241 files from imaging mode observations per day, which totals to approximately 0.4 TB of data (Observations Time Allocation Committee & Community Survey Definition Committees, 2025).

Cosmological studies with SNe Ia require extraction of SN Ia light curves from images. Difference imaging analysis (DIA) is a commonly-used technique in transient analysis (e.g., SNe Ia) that allows both transient detection and photometric measurements. Broadly, this method involves subtracting two images, one with a transient and a "reference" image without a transient, to isolate the object of interest.

Current DIA and forced photometry survey pipelines are insufficient for the volume of data expected from the Roman HLTDS survey over a 24 hour period. It would take 40 parallelized hours to make difference images (Kessler et al., 2015), resulting in a large backlog of processing, and detection inefficiencies that limit scientific discovery. There are many intermediate steps that are associated with SN Ia light curve extraction in addition to the creation of difference images; this means that the end-to-end time required to process these data from raw images to light curves will be much more that 40 hours.

*Roman* images will pose several analytical challenges: (1) the large quantity of data that will be returned on a regular basis, and (2) the spatially-dependent nature of the image properties as a function of location on the detector, which directly affects ease of (3) obtaining the required <1% flux precision on SN Ia measurements. `phrosty` is a DIA and forced photometry pipeline that addresses these challenges. We integrate the saccadic fast fourier transform (SFFT) method for difference imaging analysis (DIA, (Hu et al., 2022; Hu & Wang, 2024)), which uses a flexible delta basis function that can accommodate Roman's spatially-varying point spread function (PSF). `phrosty` uses both GPU and CPU computation; this architecture has been carefully chosen to address the challenge of processing the large quantity of survey data within a reasonable amount of time. Because Roman software is currently under development, there is no similar publicly available pipeline that takes *Roman* images as input and outputs light curves. We test our pipeline using the OpenUniverse et al. (2025) simulations; a subsequent publication will detail the accuracy and precision of our methods.

# Pipeline overview

## Pipeline steps

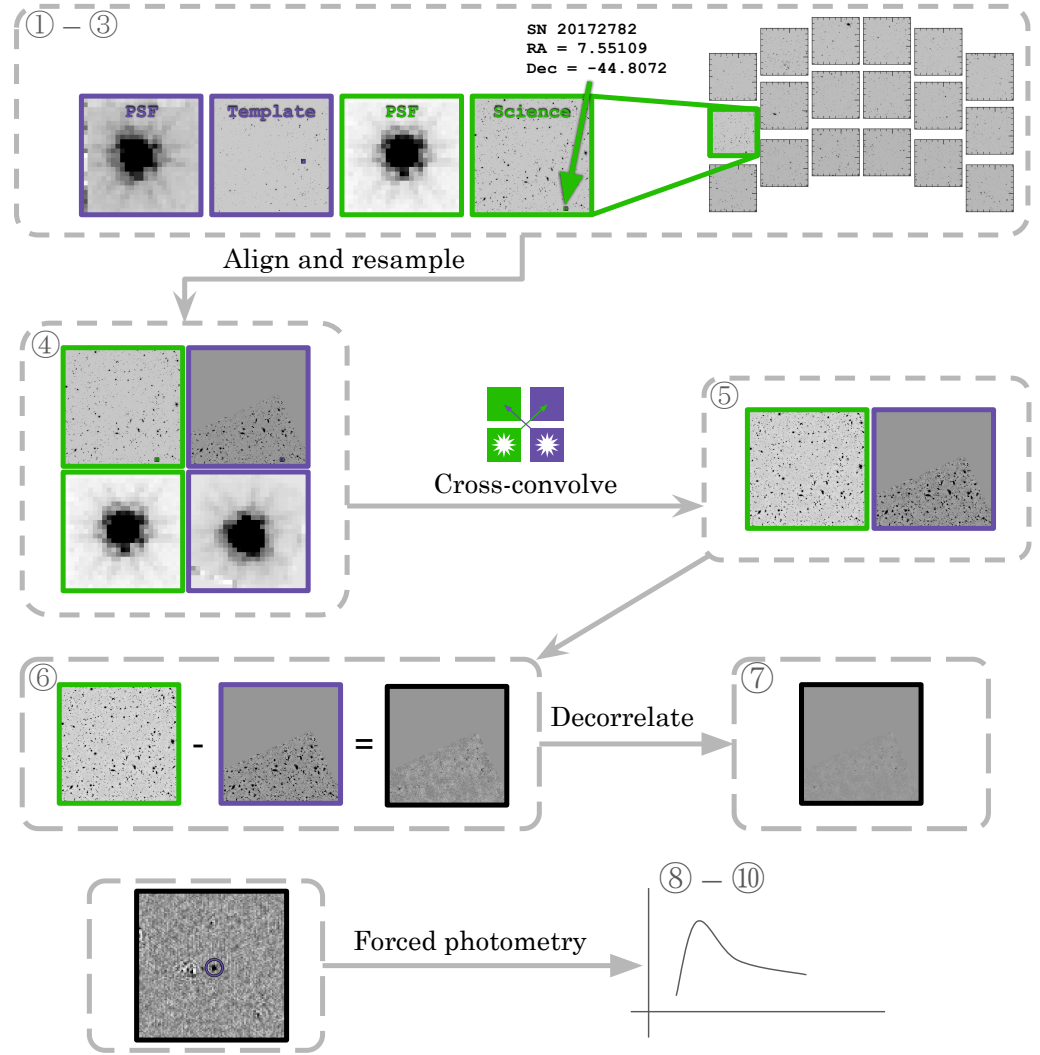

**Figure 1:** Visualization of pipeline steps using a sample image from the OpenUniverse simulations. Green outlines represent the science image, purple outlines represent the template image, and black outlines represent the difference image. The small inset green and purple squares in panel 1 represent 100 x 100 pixel cutouts.

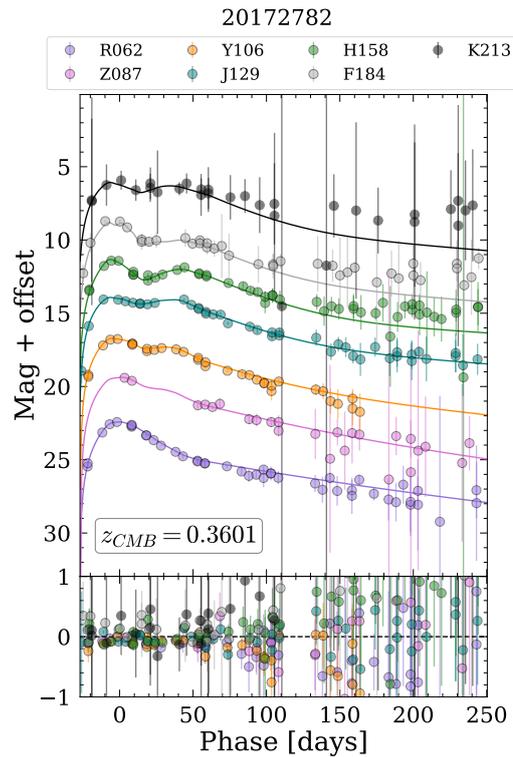

**Figure 2:** An example light curve, generated using the output of 'phrosty'. Circles indicate measured output from 'phrosty', and solid lines are the known simulated light curve from OpenUniverse.

Our pipeline steps are as follows, with "[GPU]" indicating a GPU-accelerated step (Figure 1):

1. Identify any number of appropriate science (contains SN Ia at a specified coordinate) and template images (does not contain SN Ia at the same specified coordinate) for a given SN. All input template images will be paired with all science images such that if there are $S$ science images and $T$ template images, the total number of difference images is $S \times T$.
2. The pipeline is then called from the command line interface by the user. For example,

```
SNPIT_CONFIG=phrosty/tests/phrosty_test_config.yaml python phrosty/pipeline.py \
  --oc ou2024 \
  --oid 20172782 \
  -b Y106 \
  --ic ou2024 \
  -t phrosty/tests/20172782_instances_templates_1.csv \
  -s phrosty/tests/20172782_instances_science_2.csv \
  -p 3 -w 3 \
  -v
```

`SNPIT_CONFIG` is the location of a configuration file. The inputs are: the "object collection" associated with the input image (`--oc`), the ID number assigned to the object (`--oid`), the *Roman* filter matching the data being processed (`-b`), the "image collection" associated with the input image (`--ic`), a list of template images (`-t`), a list of science images (`-s`), the number of parallel CPU processes for everything except file writing (`-p`), the number of parallel CPU process for file writing (`-w`), and a verbose flag (`-v`).

3. Subtract the sky background and generate detection masks (Bertin & Arnouts, 1996) for one pair of science and template images. Retrieve appropriate PSFs.

4. [GPU] Align the reference image, reference PSF, and reference detection mask to the science image.
5. [GPU] Cross-convolve the template PSF with the science image, and the science PSF with the template.
6. [GPU] Apply SFFT subtraction (Hu et al., 2022) to the cross-convolved images to produce a difference image.
7. [GPU] Fit and apply the decorrelation kernel to the difference image. Also apply the decorrelation kernel to the science PSF and cross-convolved science image to obtain the correct PSF for fitting, and the correct image for deriving a zero point from field stars for photometry.
8. Fit PSF from previous step to field stars in cross-convolved science image, cross-match stars to truth catalog, and record median of magnitudes for stars with $19 < m_{truth} < 21.5$ (i.e., not saturated and not noisy (Aldoroty et al., 2025)).
9. Fit the same PSF to the decorrelated difference image at the SN coordinates.
10. Steps 3–9 are repeated in serial for each science and template image pair. Light curve data are collated into a human- and machine-readable tabular *.csv file for subsequent analysis. A file is generated for each command line call. Thus, each SN has a separate file for each filter's data that is processed. These tables are sufficient to plot the measured light curve of a SN (Figure 2).

Steps 4–7 are part of the SFFT algorithm. The steps that are not GPU-accelerated, including file writing, are run in parallel processes on CPUs. The number of parallel processes is controlled by the user.

## GPU acceleration and performance

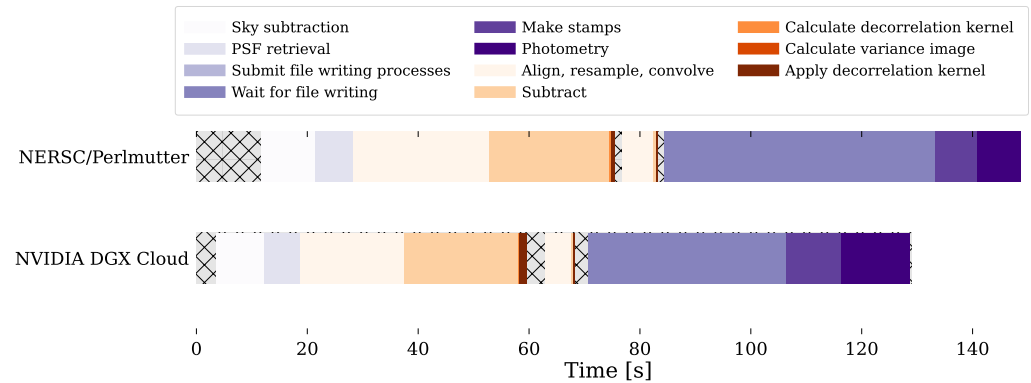

**Figure 3:** Run times gathered through NVIDIA Nsight Systems profiling for `phrosty`, run on the NVIDIA Curiosity and NERSC Perlmutter clusters. Each profile shows the end-to-end processing of two OpenUniverse images containing SN 20172782. Color bars represent functions in `phrosty`, including calls to external libraries. The grey hatched areas are other computing processes.

`phrosty` was developed in part during a hackathon hosted by NASA and Open Hackathons. During this time, `phrosty` developers, the SFFT developer (Hu et al., 2022; Hu & Wang, 2024), and NVIDIA engineers began porting more of the original SFFT package to GPU using CUDA (Nickolls et al., 2008). The final version of SFFT used by `phrosty` carries out all matrix operations on GPU, and holds all data in-memory to minimize time spent on file I/O processes. In particular, substantial effort was focused on improving image alignment and resampling code. The NVIDIA Tools Extension SDK (NVTX) is incorporated throughout the pipeline in order to enable easy code profiling, which is a unique feature compared to similar software suites.

phrosty currently runs primarily on NVIDIA A100 GPUs with 40 GB memory at the National Energy Research Scientific Computing Center's (NERSC) Perlmutter supercomputer, and has also successfully been run at the NVIDIA DGX Cluster (Figure 3), using one GPU. When subtracting full $4088 \times 4088$ images, each SN fully occupies its GPU's capacity. Thus, the number of SNe that can be run in parallel is equal to the number of GPUs available. It is possible to decrease the processed stamp size, in turn decreasing the required memory and computation time, down to a minimum size of $1000 \times 1000$ px and still obtain high-quality results.

We note that for the first time each GPU function is used (i.e., for the first image in a batch), there will be additional kernel compilation overhead time due to cupy's just-in-time (JIT) compilation. Subsequent calls are substantially faster; see Figure 3 for the run time of a two-image "light curve". Excluding file I/O, the rate-limiting steps are those that rely on external libraries: sky subtraction (Bertin & Arnouts, 1996), PSF model retrieval, and photometry (Bradley et al., 2024). All of these processes have been parallelized on CPU. Of these processes, PSF model retrieval is the most variable. In the case of cached models, the time to retrieve a PSF is dependent on file I/O. However, if a model must be generated from a process like photon shooting (OpenUniverse et al., 2025), then the time to retrieve a PSF can be much longer. Because phrosty is being developed in advance of the availability of real *Roman* images, it is compatible with simulations and both of these scenarios have been run. In the future, PSFs may also be retrieved from STPSF (Perrin et al., 2012, 2014).

## Future development

The version of phrosty presented in this work is a prototype. It will continue to be developed flexibly as *Roman* infrastructure is added and evolves. Updated versions of phrosty will be announced in subsequent publications. This includes making the pipeline fully compatible with *Roman* Science Operations Center (SOC) products. Currently, compatibility with the (OpenUniverse et al., 2025) simulations is integrated into the pipeline such that it is reliant on their metadata, and uses the FITS format versions of all files (where applicable). The final data products associated with *Roman* will not be FITS format; the project phases out the FITS file format, and replaces it with the Advanced Scientific Data Format (ASDF, (Greenfield et al., 2015)). Thus, later versions of phrosty will be fully compatible with ASDF files. phrosty will also be kept up-to-date with the latest calibration data from *Roman*, which we note may necessitate re-processing of data. This includes, but is not limited to, the PSFs discussed in Section~ and deeper co-added templates. These components of the analysis will be crucial in achieving the required $< 1\%$ flux precision from PSF photometry.

The largest barrier to phrosty's community accessibility is the computational resources it requires. Although it is fast (Figure 3), it uses nearly 40 GB of GPU memory for a single object because fast fourier transforms (FFTs) are inherently memory-intensive operations due to the large linear system it solves (Hu et al., 2022). Thus, it requires access to expensive computing resources, which are not always available to the entire astronomical community. Although this hardware becomes more readily available over time as technological advancements continue, and we do not expect phrosty to ever require more resources than it demands in its current status, one of our goals for future development is to reduce phrosty's memory consumption.

## Acknowledgements

L. A. thanks Megan Sosey for the discussion about *Roman* HLTDS data volume.

Funding for the Roman Supernova Project Infrastructure Team has been provided by NASA under contract to 80NSSC24M0023. This work is also supported by NASA under award number 80GSFC24M0006. M. T. was funded by NASA under JPL Contract Task 70-711320, "Maximizing Science Exploitation of Simulated Cosmological Survey Data Across Surveys''.

This research used resources of the National Energy Research Scientific Computing Center, which is supported by the Office of Science of the U.S. Department of Energy using award number HEP-ERCAP32751.

This work was completed in part at the NASA Open Hackathon, part of the Open Hackathons program. The authors would like to acknowledge OpenACC-Standard.org for their support. The authors acknowledge and thank NVIDIA for their contributions.

# References

Akeson, R., Armus, L., Bachelet, E., Bailey, V., Bartusek, L., Bellini, A., Benford, D., Bennett, D., Bhattacharya, A., Bohlin, R., Boyer, M., Bozza, V., Bryden, G., Calchi Novati, S., Carpenter, K., Casertano, S., Choi, A., Content, D., Dayal, P., … Zimmerman, N. (2019). The Wide Field Infrared Survey Telescope: 100 Hubbles for the 2020s. *arXiv e-Prints*, arXiv:1902.05569. https://doi.org/10.48550/arXiv.1902.05569

Aldoroty, L., Scolnic, D., Kannawadi, A., Knop, R., Rose, B., Hounsell, R., & Troxel, M. (2025). Initial Characterization of Stellar Photometry of Roman images from the OpenUniverse Simulations. *arXiv e-Prints*, arXiv:2506.04332. https://doi.org/10.48550/arXiv.2506.04332

Bertin, E., & Arnouts, S. (1996). SExtractor: Software for source extraction. *Astronomy and Astrophysics Supplement*, *117*, 393–404. https://doi.org/10.1051/aas:1996164

Bradley, L., Sipőcz, B., Robitaille, T., Tollerud, E., Vinícius, Z., Deil, C., Barbary, K., Wilson, T. J., Busko, I., Donath, A., Günther, H. M., Cara, M., Lim, P. L., Meßlinger, S., Burnett, Z., Conseil, S., Droettboom, M., Bostroem, A., Bray, E. M., … Perren, G. (2024). *Astropy/photutils: 1.13.0* (Version 1.13.0). Zenodo. https://doi.org/10.5281/zenodo.12585239

Greenfield, P., Droettboom, M., & Bray, E. (2015). ASDF: A new data format for astronomy. *Astronomy and Computing*, *12*, 240–251. https://doi.org/10.1016/j.ascom.2015.06.004

Hu, L., & Wang, L. (2024). Differencing and Coadding JWST Images with Matched Point-spread Function. *The Astronomical Journal*, *167*(5), 231. https://doi.org/10.3847/1538-3881/ad36cb

Hu, L., Wang, L., Chen, X., & Yang, J. (2022). Image Subtraction in Fourier Space. *The Astrophysical Journal*, *936*(2), 157. https://doi.org/10.3847/1538-4357/ac7394

Kessler, R., Marriner, J., Childress, M., Covarrubias, R., D'Andrea, C. B., Finley, D. A., Fischer, J., Foley, R. J., Goldstein, D., Gupta, R. R., Kuehn, K., Marcha, M., Nichol, R. C., Papadopoulos, A., Sako, M., Scolnic, D., Smith, M., Sullivan, M., Wester, W., … DES Collaboration. (2015). The Difference Imaging Pipeline for the Transient Search in the Dark Energy Survey. *150*(6), 172. https://doi.org/10.1088/0004-6256/150/6/172

Nickolls, J., Buck, I., Garland, M., & Skadron, K. (2008). Scalable parallel programming with CUDA: Is CUDA the parallel programming model that application developers have been waiting for? *Queue*, *6*(2), 40–53. https://doi.org/10.1145/1365490.1365500

Observations Time Allocation Committee, R., & Community Survey Definition Committees, C. (2025). Roman Observations Time Allocation Committee: Final Report and Recommendations. *arXiv e-Prints*, arXiv:2505.10574. https://doi.org/10.48550/arXiv.2505.10574

OpenUniverse, The LSST Dark Energy Science Collaboration, The Roman HLIS Project Infrastructure Team, The Roman RAPID Project Infrastructure Team, The Roman Supernova Cosmology Project Infrastructure Team, Alarcon, A., Aldoroty, L., Beltz-Mohrmann, G., Bera, A., Blazek, J., Bogart, J., Braeunlich, G., Broughton, A., Cao, K., Chiang, J., Chisari, N. E., Desai, V., Fang, Y., Galbany, L., … Zhang, T. (2025). OpenUniverse2024: A shared, simulated view of the sky for the next generation of cosmological surveys. *arXiv e-Prints*, arXiv:2501.05632. https://doi.org/10.48550/arXiv.2501.05632

Perrin, M. D., Sivaramakrishnan, A., Lajoie, C.-P., Elliott, E., Pueyo, L., Ravindranath, S., & Albert, Loïc. (2014). Updated point spread function simulations for JWST with WebbPSF. In J. M. Oschmann Jr., M. Clampin, G. G. Fazio, & H. A. MacEwen (Eds.), *Space telescopes and instrumentation 2014: Optical, infrared, and millimeter wave* (Vol. 9143, p. 91433X). https://doi.org/10.1117/12.2056689

Perrin, M. D., Soummer, R., Elliott, E. M., Lallo, M. D., & Sivaramakrishnan, A. (2012). Simulating point spread functions for the James Webb Space Telescope with WebbPSF. In M. C. Clampin, G. G. Fazio, H. A. MacEwen, & J. M. Oschmann Jr. (Eds.), *Space telescopes and instrumentation 2012: Optical, infrared, and millimeter wave* (Vol. 8442, p. 84423D). https://doi.org/10.1117/12.925230

Spergel, D., Gehrels, N., Baltay, C., Bennett, D., Breckinridge, J., Donahue, M., Dressler, A., Gaudi, B. S., Greene, T., Guyon, O., Hirata, C., Kalirai, J., Kasdin, N. J., Macintosh, B., Moos, W., Perlmutter, S., Postman, M., Rauscher, B., Rhodes, J., … Zhao, F. (2015). Wide-Field InfrarRed Survey Telescope-Astrophysics Focused Telescope Assets WFIRST-AFTA 2015 Report. *arXiv e-Prints*, arXiv:1503.03757. https://doi.org/10.48550/arXiv.1503.03757

Spergel, D., Gehrels, N., Breckinridge, J., Donahue, M., Dressler, A., Gaudi, B. S., Greene, T., Guyon, O., Hirata, C., Kalirai, J., Kasdin, N. J., Moos, W., Perlmutter, S., Postman, M., Rauscher, B., Rhodes, J., Wang, Y., Weinberg, D., Centrella, J., … Shaklan, S. (2013). WFIRST-2.4: What Every Astronomer Should Know. *arXiv e-Prints*, arXiv:1305.5425. https://doi.org/10.48550/arXiv.1305.5425